\documentclass{PoS}
\pdfoutput=1
\usepackage{rotating,booktabs,amsmath, multicol}
\usepackage[font=footnotesize,labelfont=bf,labelsep=period]{caption}
\usepackage[numbers,sort&compress]{natbib}
\usepackage{natbibspacing}
\newcommand{\comment}[1]{}

\providecommand{\abs}[1]{\lvert#1\rvert}    

\title{$B$-physics with dynamical domain-wall light quarks and relativistic $b$-quarks}
\ShortTitle{$B$-physics with dynamical domain-wall light quarks and relativistic $b$-quarks}

\author{Ruth S.~Van de Water and \speaker{Oliver Witzel}\\
        Brookhaven National Laboratory, Department of Physics, Upton, NY 11973, USA\\
        E-mail: \email{witzel@quark.phy.bnl.gov}}

\abstract{We report on our progress in calculating the $B$-meson decay constants and $B^0$-$\bar B^0$ mixing parameters using domain-wall light quarks and relativistic $b$-quarks. We present our computational method and show some preliminary results obtained on the coarser ($a \approx 0.11$fm) $24^3$ lattices. This work is presented on behalf of the RBC and UKQCD collaborations.}

\FullConference{The XXVIII International Symposium on Lattice Field Theory, Lattice2010\\
		June 14-19, 2010\\
		Villasimius, Italy}

\begin{document}

\section{Introduction}\vspace{-2mm}
The study of $B$-meson physics on the lattice is of special phenomenological interest because it allows one to obtain constraints on the CKM unitarity triangle. In the standard global unitarity triangle fit, the apex of the CKM unitarity triangle is constrained using lattice input for neutral $B$-meson mixing. Experimentally, $B_q$--$\bar B_q$ mixing is well measured in terms of mass differences (oscillation frequencies) $\Delta m_q$; in the standard model it is parameterized by \cite{Buras:1990fn}
\begin{align} 
\Delta m_q = \frac{G_F^2m^2_W}{6\pi^2} \eta_B S_0 m_{B_q}{f_{B_q}^2B_{B_q}} \abs{V_{tq}^*V_{tb}}^2,
\end{align}
where the index $q$ denotes either a $d$- or a $s$-quark, $m_{B_q}$ is the mass of the $B_q$-meson, and  $V_{tq}^*$ and $V_{tb}$ are CKM matrix elements. The Inami-Lim function, $S_0$ \cite{Inami:1980fz}, and the QCD coefficient, $\eta_B$ \cite{Buras:1990fn}, can be computed perturbatively, whereas the non-perturbative input is given by $f_q^2 B_{B_q}$: the leptonic decay constant $f_{B_q}$ and the $B$-meson bag parameter $B_{B_q}$. Computing the ratio $\Delta m_s$/$\Delta m_d$ is particularly advantageous because statistical and systematic uncertainties largely cancel and, moreover, the ratio of CKM matrix elements becomes accessible
\begin{align}
\frac{\Delta m_s}{\Delta m_d} = \frac{m_{B_s}}{m_{B_d}}\,{\xi^2} \, \frac{\abs{V_{ts}}^2}{\abs{V_{td}}^2}.
\end{align}
The non-perturbative input is now solely contained in the $SU(3)$-breaking ratio 
\begin{align}
\xi &= \frac{f_{B_s}\sqrt{B_{B_s}}}{f_{B_d}\sqrt{B_{B_d}}}.
\end{align}

An alternative way of constraining the CKM triangle has been proposed by Lunghi and Soni \cite{Lunghi:2009ke} that uses the CKM matrix elements $V_{ub}$ or $V_{cb}$, thereby avoiding the tension between inclusive and exclusive determinations of both $V_{ub}$ (3$\sigma$) and $V_{cb}$ (2$\sigma$). This alternative method, however, requires precise knowledge of the decay constant $f_{B}$ as well as $BR(B \to \tau \nu)$ and $\Delta m_s$.  Moreover, Lunghi and Soni point out \cite{Lunghi:2008aa,Lunghi:2010gv} that already the current data may show signs of physics beyond the standard model manifesting themselves as deviations of experimental values for $\sin(2\beta)$ (3.3$\sigma$) and $BR(B\to\tau l \nu)$ (2.8$\sigma$) from the Standard Model. They argue that the most likely sources for new physics are in $B_q$ mixing and $\sin(2\beta)$.

Therefore, it is timely for the lattice community to determine these parameters precisely and prove by using statistically and systematically independent setups that all sources of errors are well under control. Currently, there are three determinations for $\xi$ using 2+1-flavor gauge field configurations: an exploratory study by the RBC-UKQCD collaboration \cite{Albertus:2010nm} with non-competitive errors and two precise determinations by Fermilab-MILC \cite{ToddEvans:2008,Evans:2009du} and HPQCD \cite{Gamiz:2009ku}, which however both rely on the same set of MILC gauge field configurations. The latter two collaborations also use the same configurations to obatin the decay constants $f_{B_d}$ and $f_{B_s}$ and find good agreement. Here we present the first results of our project to determine $B$-meson parameters using domain-wall light quarks and relativistic $b$-quarks.

\section{Computational setup}\vspace{-2mm}
Our project is based on using the RBC-UKQCD 2+1-flavor domain wall lattices with the Iwasaki gauge action for the gluons. Our first results presented in these proceedings are obtained on the coarser $a\approx 0.11$fm lattices with a spatial volume of $24^3$ in lattice units \cite{Allton:2008pn}, whereas the full project will also utilize the finer $a\approx 0.08$fm lattices ($32^3$) \cite{Aoki:2010dy} (for details see Tab.~\ref{Tab-lattices}). 
\begin{table}[hbt]
\centering
\begin{tabular}{ccllcrc}\toprule
  &         &         &         &            & approx.~~&\# time\\
L & $a$(fm) & ~~$m_l$ & ~~$m_s$ & $m_\pi$(MeV)& \# configs.&sources\\ \midrule
24 & $\approx$ 0.11 &  0.005 & 0.040 & 331 & 1640~~~~&1\\
24 & $\approx$ 0.11 &  0.010 & 0.040 & 419 & 1420~~~~&1\\
24 & $\approx$ 0.11 &  0.020 & 0.040 & 558 & 350~~~~&8\\
\midrule
32 & $\approx$ 0.08 &  0.004 & 0.030 & 307 & 600~~~~&1\\
32 & $\approx$ 0.08 &  0.006 & 0.030 & 366 & 900~~~~&1\\
32 & $\approx$ 0.08 &  0.008 & 0.030 & 418 & 550~~~~&1\\ \bottomrule
\end{tabular}
\caption{Overview of the gauge field ensembles to be used for this project.}
\label{Tab-lattices}
\end{table}

\begin{figure}[htb]
\centering
\includegraphics[scale=0.4,clip]{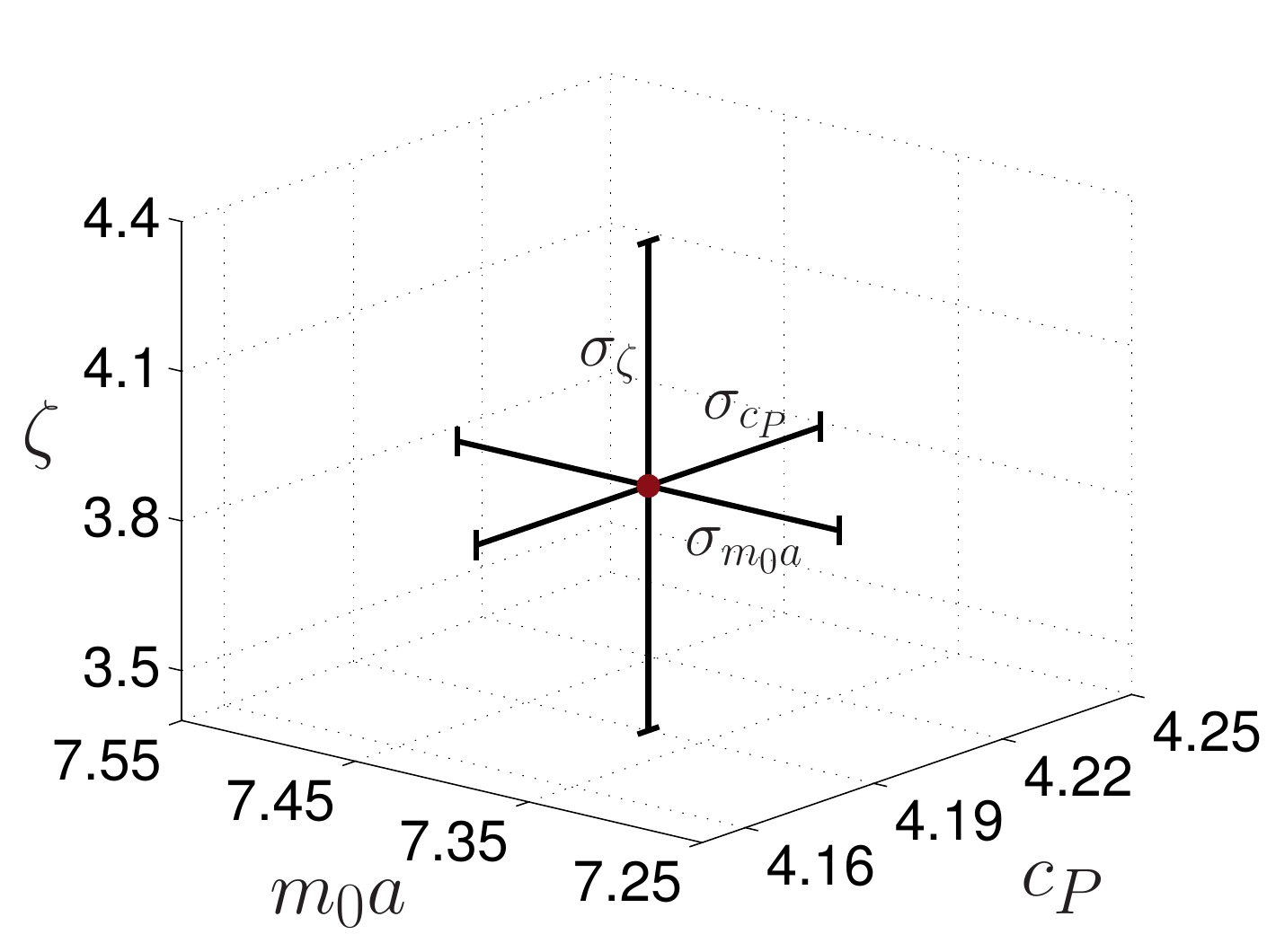}
\caption{Initial guess for the RHQ parameters and their uncertainties.}
\label{Fig-RHQparameters}
\end{figure}

For the heavy quarks we use a relativistic formulation derived from the ``Fermilab action''  \cite{ElKhadra:1996mp} in which we tune the three relevant parameters of the action non-perturbatively. As demonstraged by Christ, Li and Lin, this requires only one additional experimental input compared to the formulation of Fermilab \cite{Christ:2006us}. The relativistic heavy quark (RHQ) action is given by
\begin{align}
S = \sum_{n,n^\prime}\bar\Psi_n \left\{\! m_0+ \gamma_0 D_0 -\! \frac{a D_0^2}{2} + \zeta\left[\vec \gamma \cdot \vec D- \frac{a\left(\vec D \right)^2}{2}\right]\!- a \sum_{\mu\nu} \frac{i c_P}{4}\sigma_{\mu\nu}F_{\mu\nu}\!\right\}_{\!\!n,n^\prime}\!\!\!\!\Psi_{n^\prime},
\end{align}
where the covariant derivative is denoted by $D$, $F_{\mu\nu}$ is the field strength tensor and we need to tune the three parameters $m_0a$, $c_P$ and $\zeta$. Exploratory studies on how to tune these parameters have been performed by Li and Peng \cite{Li:2008kb,Peng:2009,Peng:2010}. The idea is to use experimental values for the spin averaged meson mass ($\overline{m} = (m_{B_s}+3m_{B_s^*})/4$) and the hyperfine splitting ($\Delta_m = m_{B^*_s}-m_{B_s}$) as inputs together with the constraint from the dispersion relation that the rest mass $m_1$ equal the kinetic mass $m_2$. These three quantities are computed for a set of seven trial parameters determined  by making an initial guess for $m_0a$, $c_P$, and $\zeta$ and then varying it by a chosen uncertainty $\pm \sigma_{\{m_0a,c_P,\zeta\}}$:
\begin{align}
\left[\!\!\begin{array}{c} m_0 a\\ c_P\\ \zeta\\ \end{array}\right],
\left[\!\!\begin{array}{c}m_0 a -\sigma_{m_0 a}\\ c_P\\\zeta \\ \end{array}\!\!\right],\;
\left[\!\!\begin{array}{c}m_0 a +\sigma_{m_0 a}\\ c_P\\\zeta \\ \end{array}\!\!\right],\;
\left[\!\!\begin{array}{c}m_0 a\\c_P-\sigma_{c_P}\\ \zeta\\ \end{array}\!\!\right],\;
\left[\!\!\begin{array}{c}m_0 a\\c_P+\sigma_{c_P}\\ \zeta\\ \end{array}\!\!\right],\;
\left[\!\!\begin{array}{c}m_0 a\\ c_P\\ \zeta-\sigma_{\zeta}\\ \end{array}\!\!\right],\;
\left[\!\!\begin{array}{c}m_0 a\\ c_P\\ \zeta+\sigma_{\zeta}\\ \end{array}\!\!\right]
\label{Eq-Inputs}
\end{align}
(see Fig.~\ref{Fig-RHQparameters}). We iterate over the parameters $\{m_0a,\,c_P,\, \zeta\}$ until we determine the values that reproduce the known experimental measurements for $\overline{m},\, \Delta_m,\, m_1/m_2$:
\begin{align}
&\left[\begin{array}{c} m_0a\\ c_P \\ \zeta \end{array}\right]^\text{RHQ} = J^{-1} \times\left(\left[\begin{array}{c}\overline{m}\\ \Delta_m\\ \frac{m_1}{m_2}\end{array}\right]^\text{PDG} - A\right) \label{Eq-RHQdetermination}
\intertext{with}
\displaystyle &J = \left[\frac{Y_3- Y_2}{2\sigma_{m_0a}},\,\frac{Y_5-Y_4}{2\sigma_{c_P}},\,\frac{Y_7 - Y_6}{2\sigma_\zeta}\right]
\qquad\text{and}\quad
A = Y_1 - J\times \left[m_0a,\, c_P,\, \zeta\right]^t. \label{Eq-JAmatrices}
\end{align}
In (\ref{Eq-JAmatrices}) we use the vectors $Y_i$ as shorthand notation for $[\overline{m},\,\Delta_m,\,m_1/m_2]^t_i$ obtained for the input parameters as given in Eq.~(\ref{Eq-Inputs}), labeled 1-7 from left to right. Eq.~(\ref{Eq-RHQdetermination}) assumes a linear dependence of the meson masses on the parameters of the action; therefore we must be in a linear regime to reliably extract $\{m_0a,\,c_P,\,\zeta\}$. We stop the iteration once the tuned values lie within our variation range.

In order to finally compute $B$-meson decay constants and mixing parameters as precisely as possible we repeated the original tuning presented at Lattice 2008 \cite{Li:2008kb} with increased statistics and optimized smeared wavefunction parameters used for generating the heavy quark propagators. Moreover, performing the tuning on the same configurations we intend to use for computing weak matrix elements such as $f_B$ will allow for an improved error analysis in which the correlations among the three RHQ parameters can be fully taken into account. 

\begin{figure}[htb]
\centering
\includegraphics[scale=0.4,clip]{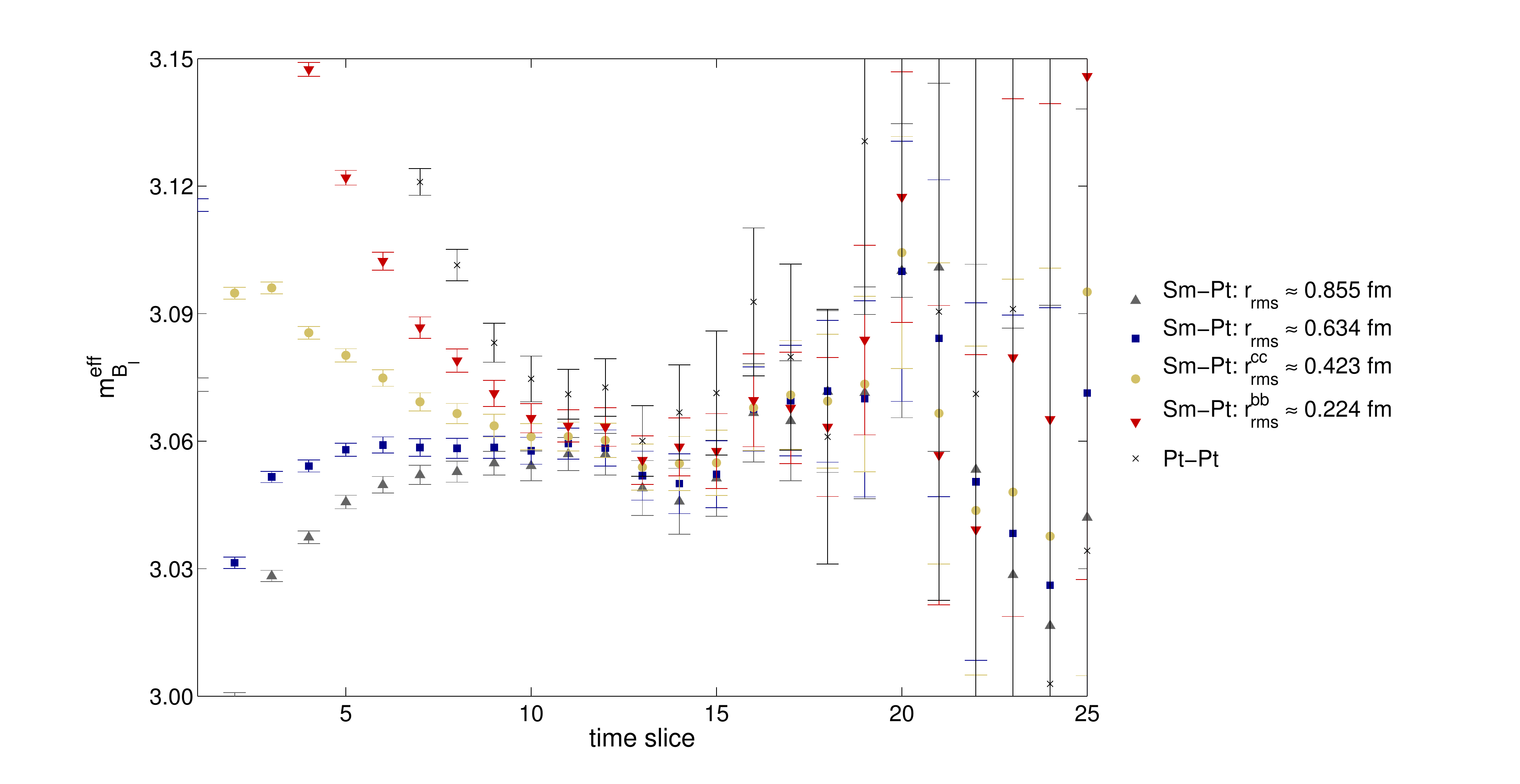}
\caption{$B_l$-meson effective masses for different $b$-quark spatial wavefunctions; in each case the light quark has a point source.}
\label{Fig-SmearingTest}
\end{figure}

Our method for computing the $\Delta B =2$ four-quark operators requires the light quarks to be generated with a point source and sink, but for the heavy quarks we are free to explore different smearing choices. In addition to point sources and sinks, we tried Gaussian smeared sources/sinks and also varied the radius of the Gaussian smearing. As a first guess we chose the radius of the Gaussian source to be the rms radius of the $b\bar b$- and $c \bar c$ states \cite{Menscher:2005kj}. Later we extended the radius and found that $r_\text{rms} =0.634$fm gives the best signal as can be seen in Fig.~\ref{Fig-SmearingTest}.

Using this setup we obtain the RHQ parameters on the three $24^3$ ensembles as given in Table \ref{Tab-TunedRHQparameters}. In contrast to earlier work \cite{Li:2008kb} the determination uses only quantities from the heavy-light system. We expect these values to be close to our final determination. Moreover, we observe that within statistical uncertainties there is no dependence on the light sea quark mass ($m_\text{sea}^l$).

\begin{table}[htb]
\centering
\begin{tabular}{clll}\toprule
 $m_\text{sea}^l$ & $m_0a$ & $c_P$ & $\zeta$ \\ \midrule
 0.005 & 8.41(9)& 5.7(2) & 3.1(2)\\
 0.010 & 8.4(1) & 5.8(2) & 3.1(1)\\
 0.020 & 8.4(1) & 5.6(2) & 3.1(1)\\\bottomrule
\end{tabular}
\caption{Preliminary determination of the RHQ parameters on the three $24^3$ ensembles with $a \approx 0.11$fm.}
\label{Tab-TunedRHQparameters}
\end{table}

\section{Computation of decay constants}
\begin{figure}[htb]
\centering
\includegraphics[scale=0.36,clip]{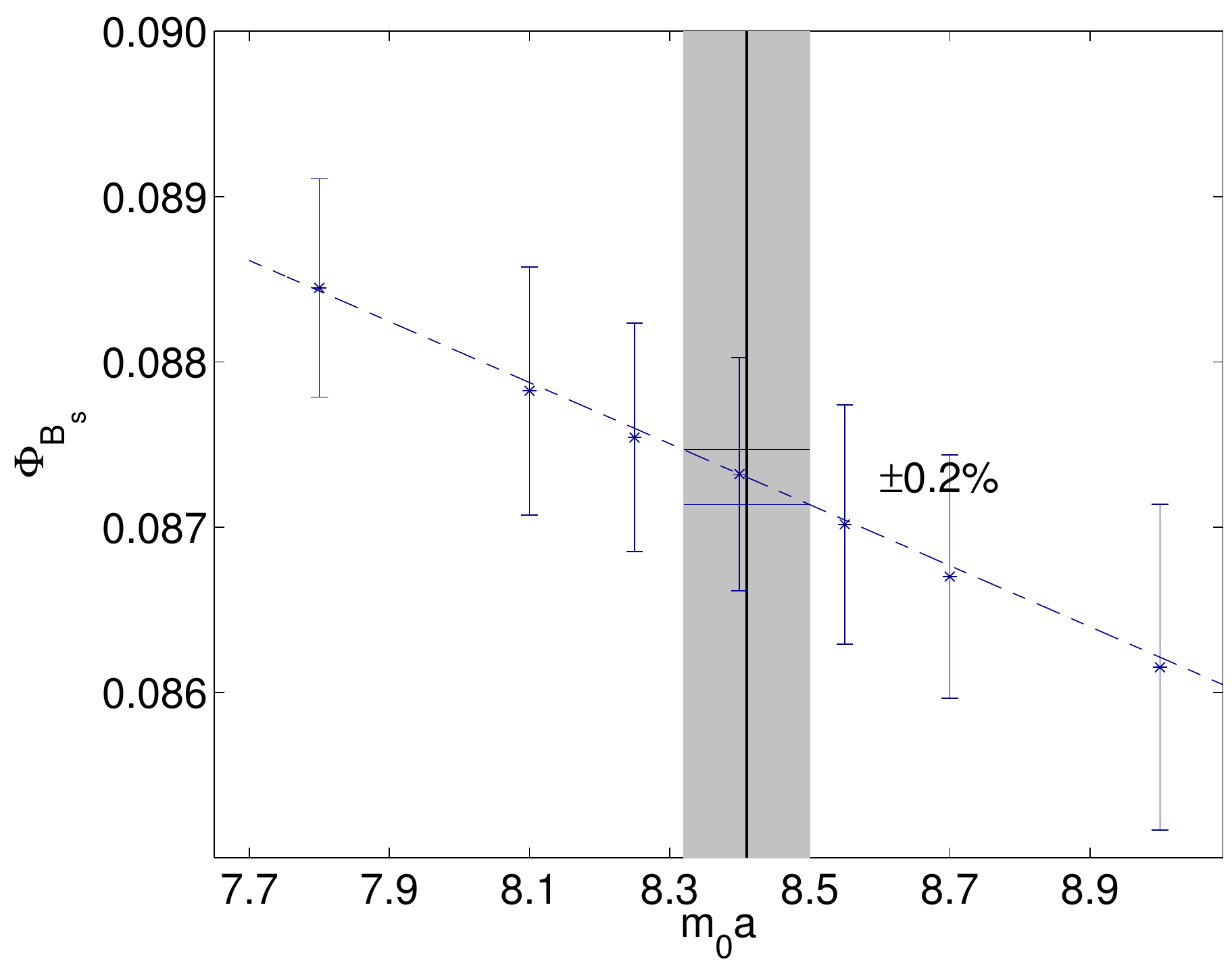}
\includegraphics[scale=0.36,clip]{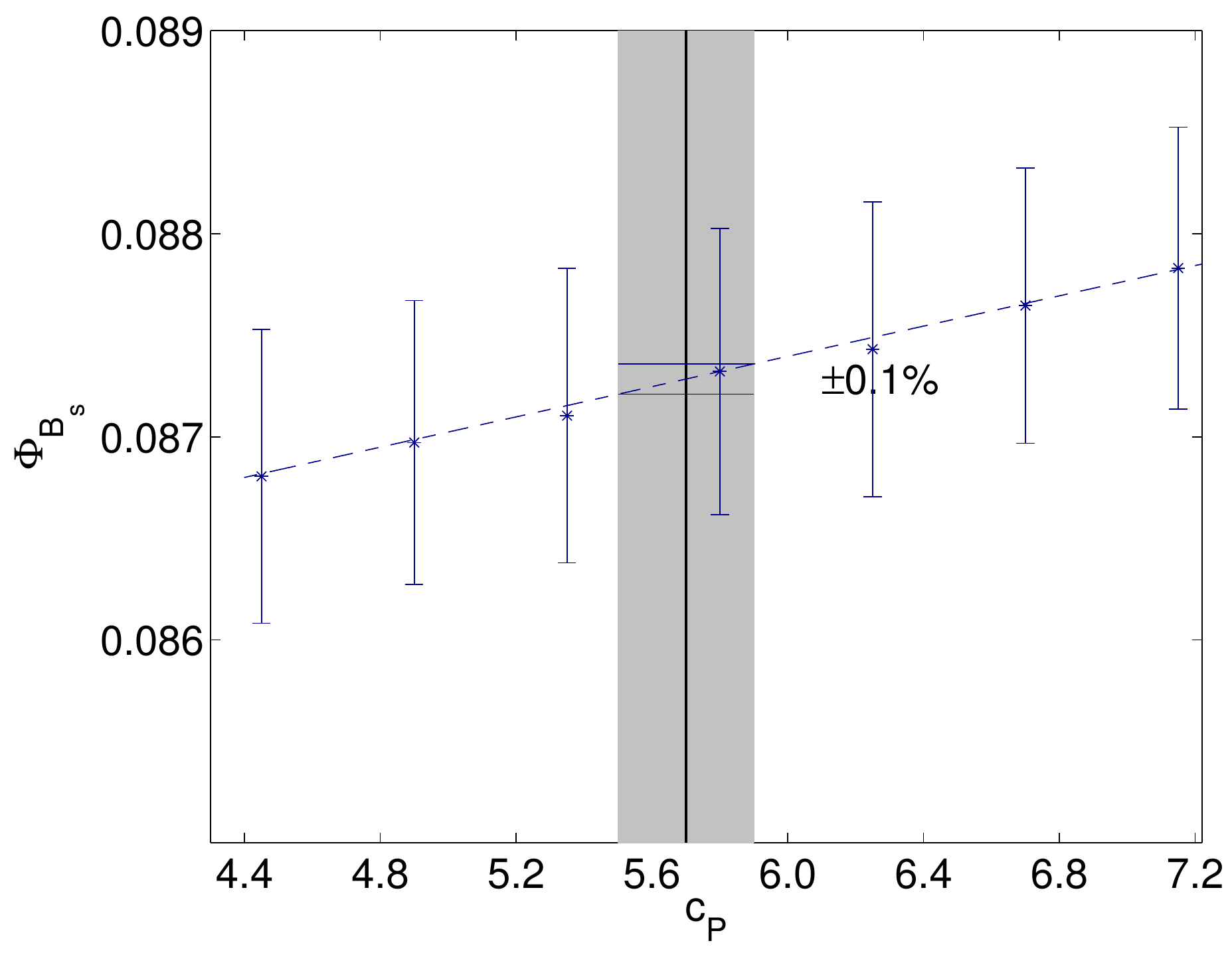}
\includegraphics[scale=0.36,clip]{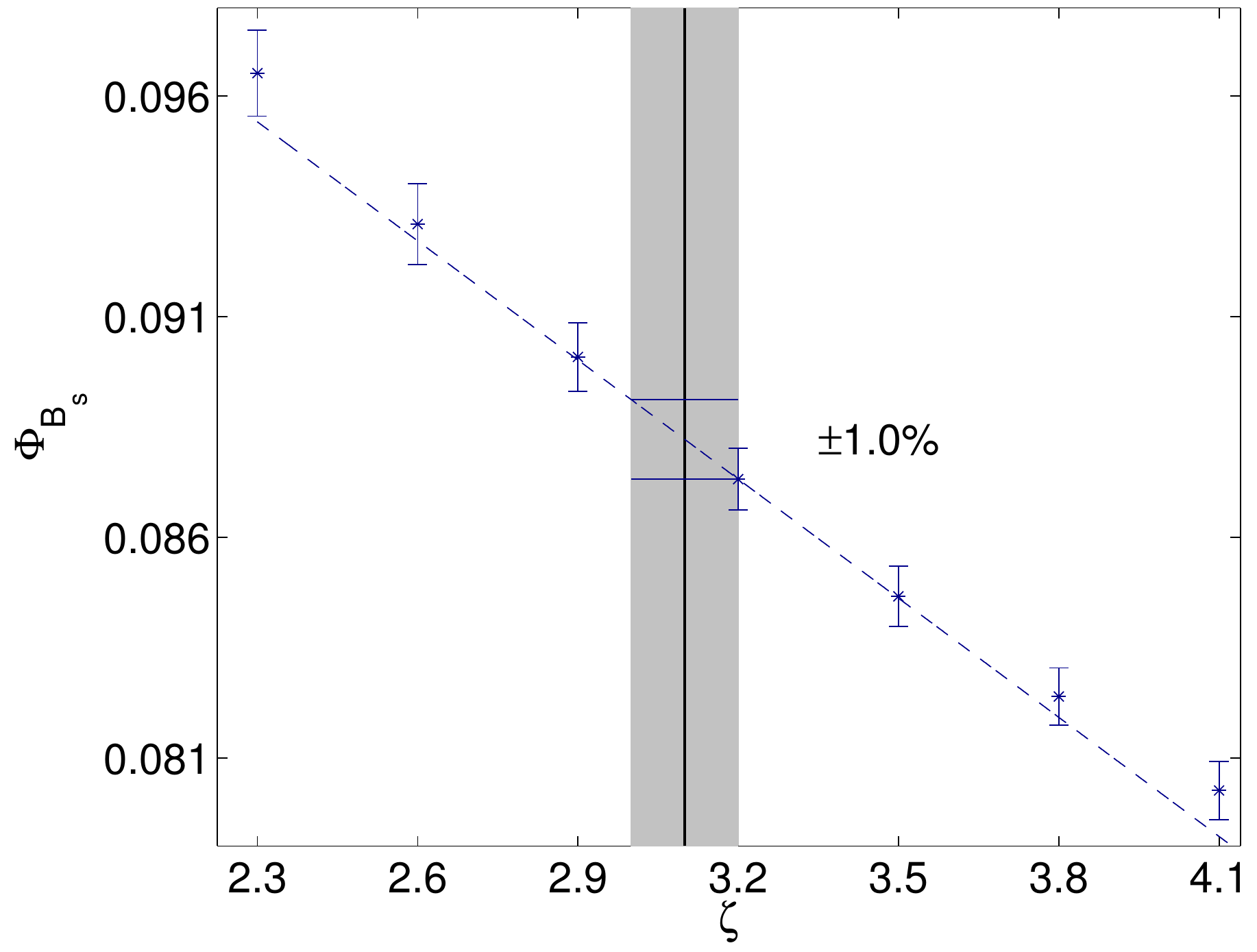}
\caption{Dependence of $\Phi_{B_s}$ on the three RHQ parameters $\{m_0a,\,c_P,\,\zeta\}$; results are shown for the lightest sea quark mass $m_\text{sea}^l=0.005$ on the $24^3$ ensembles.}
\label{Fig-PhiRHQ}
\end{figure}

\vspace{-2mm}
Using these newly determined RHQ parameters we compute as a first non-trivial test the decay constants of the unitary ($B_l$) and strange ($B_s$) mesons via the relation
\begin{align}
f_{B_q} = Z_\Phi\; \Phi_{B_q}\; a^{-3/2} / \sqrt{m_{B_q}},
\end{align}
where $\Phi_{B}$ is the lattice decay amplitude, $Z_\Phi$ is the renormalization factor. For the $B_s$ meson (domain-wall valence quark has mass of physical $s$-quark) we generated data for the set of seven different RHQ input parameters. Therefore we can use similar equations to (\ref{Eq-RHQdetermination}) and (\ref{Eq-JAmatrices}) in order to determine the decay amplitude $\Phi_{B_s}$ at the tuned RHQ parameters:
\begin{align}
&\Phi^\text{RHQ} = J_\Phi^{(1\times 3)} \times \left[\begin{array}{c} m_0a\\ c_P \\ \zeta \end{array}\right]^\text{RHQ}+ A_\Phi \label{Eq-RHQprediction}
\intertext{with}
&\displaystyle J_\Phi = \left[\frac{\Phi_3-\Phi_2}{2\sigma_{m_0a}},\, \frac{\Phi_5-\Phi_4}{2\sigma_{c_P}},\, \frac{\Phi_7-\Phi_6}{2\sigma_{\zeta}}\right] \qquad\text{and}\qquad A_\Phi = \Phi_1 - J_\Phi \times \left[m_0a,\, c_P,\,\zeta \right]^t.
\end{align}
In Fig.~\ref{Fig-PhiRHQ} we show the results for $\Phi_{B_s}$ obtained on the ensemble with $m_\text{sea}^l=0.005$. To test the linearity with respect to the input parameters $\{m_0a,\,c_P,\,\zeta\}$  we used three different sets of seven RHQ parameters all centered around the same point. The vertical black line with the gray error band indicates the tuned value of $m_0a$, $c_P$ or $\zeta$ and allows for a simple estimate of the error in $\Phi_{B_s}$ due to the uncertainty in each of the three parameters. These plots show that the effect of the uncertainty in $m_0a$ and $c_P$ is negligible, whereas $\zeta$ contributes an error of about 1\%. Moreover, we see that $\Phi_{B_s}$ depends linearly on $m_0a$, $c_P$ and $\zeta$ in the range of interest.

\vspace{-1mm}
Alternatively, one may simply use the tuned RHQ values to compute the needed correlation functions. We followed this procedure to calculate the $B_l$ meson (domain-wall valence quark mass equals the light sea quark mass) the decay amplitude $\Phi_{B_l}$.

\begin{figure}[htb]
\flushright
\begin{minipage}{0.45\textwidth}
\centering
\includegraphics[scale=0.33]{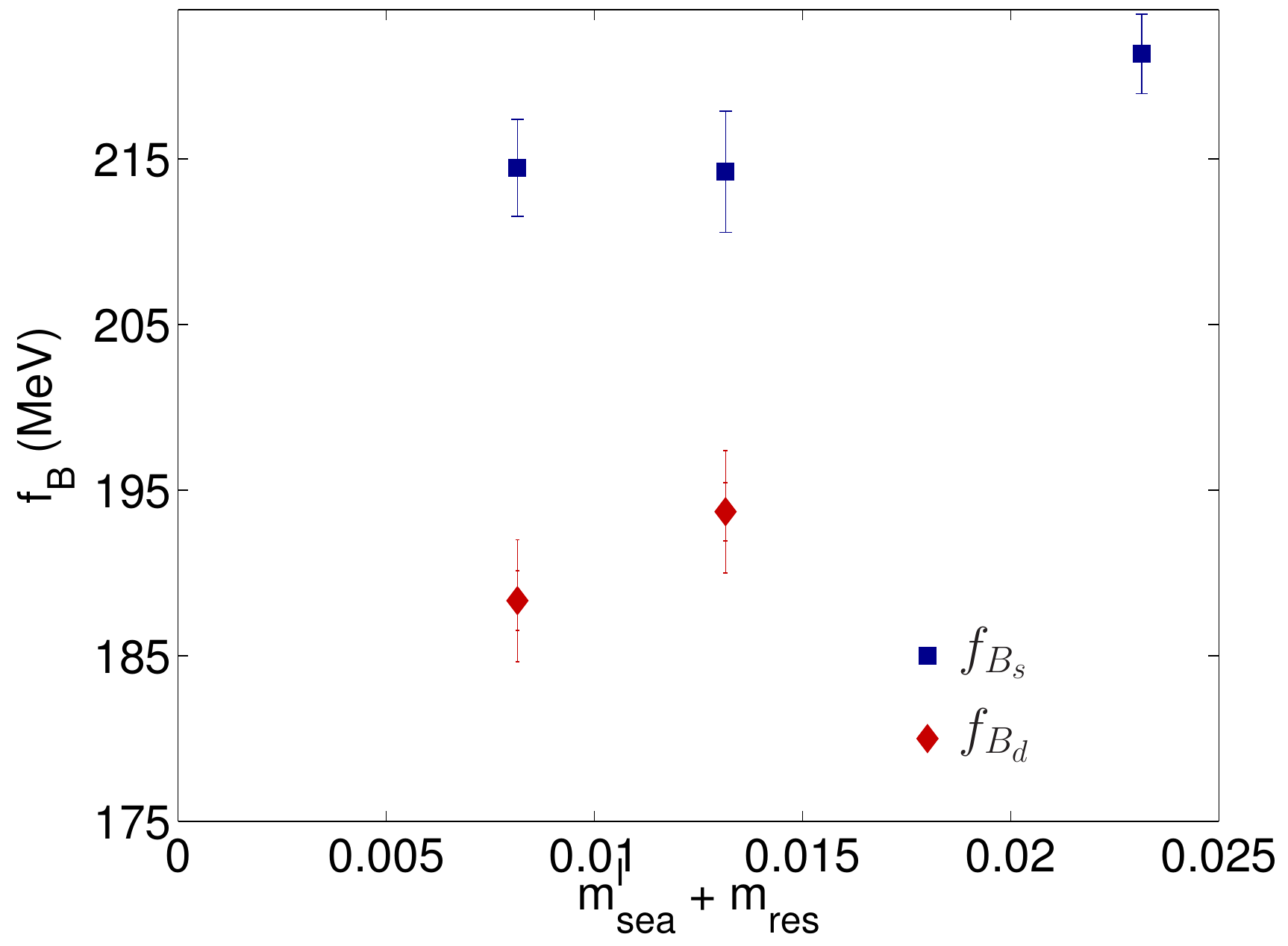}
\caption{Preliminary decay constants $f_{B_l}$ and $f_{B_s}$ computed on our three $24^3$ ensembles.}
\label{Fig-fB}
\end{minipage}
\end{figure}

Atfer renormalizing $\Phi_B$ multiplicatively at 1-loop \cite{Yamada:2004ri} we obtain the decay constants which are given in Tab.~\ref{Tab-fB} and shown in Fig.~\ref{Fig-fB}. As mentioned above, the values for $f_{B_l}$ are obtained by simulating directly at the tuned RHQ parameters, whereas  the $f_{B_s}$ values are extracted using Eq.~(\ref{Eq-RHQprediction}). The statistical errors on $f_{B_s}$ are therefore larger because they take into account the statistical uncertainties in the three RHQ parameters. For a better comparison, we increase the errors in $f_{B_l}$  in the Fig.~\ref{Fig-fB} by the error due to the statistical uncertainty in the RHQ parameters estimated in Fig.~\ref{Fig-PhiRHQ}. Moreover, we emphasize that this computation is performed without $O(a)$ improvement.

\section{Conclusion}\vspace{-2mm}
We presented our first results computing $B$-meson decay constants using domain-wall light and relativistic $b$-quarks with all parameters of the RHQ action tuned non-perturbatively. Currently, our results still need to be ${\cal O}(a)$ improved. For $f_{B_s}$ we expect only a mild chiral extrapolation but of course need to perform an extrapolation to the continuum and estimate other systematic errors.  Despite these caveats the small statistical errors and the fact that our central values lie in the same ballpark as results of other collaborations indicates the promise of our method.

\clearpage
\begin{table}[htb]
\begin{minipage}{0.45\textwidth}
\centering
\begin{tabular}{ccc}\toprule
 $m_{sea}^l$ & $f_{B_l}$(MeV) & $f_{B_s}$(Mev) \\\midrule
 0.005 &  188(2) & 215(3) \\ 
 0.010 &  194(2) & 214(4)\\
 0.020 &  ---    & 221(2)\\\bottomrule
\end{tabular}
\caption{Preliminary results for the decay constants using 1-loop multiplicative renormalization, but without $O(a)$ improvement of the axial-current operator.}
\label{Tab-fB}
\end{minipage}
\end{table}
\vspace{-50mm}

\vspace{-6mm}
\section*{Acknowledgments}\vspace{-3mm}
We are thankful to all the members of the RBC and UKQCD 
collaborations.  
Numerical computations for this work utilized USQCD resources 
and were performed on 
the kaon and jpsi clusters at FNAL, in part funded by the 
Office of Science of the U.S.~Department of Energy.
This manuscript has been authored by 
an employee of Brookhaven Science Associates, LLC 
under Contract No.~DE-AC02-98CH10886 with the 
U.S.~Department of Energy.\vspace{-3mm}

\bibliography{B_meson}
\bibliographystyle{apsrev4-1}

\end{document}